Reflection high-energy electron diffraction and scanning tunneling microscopy study of InP(001) surface reconstructions

V.P. LaBella, Z. Ding, D.W. Bullock, C. Emery, and P.M. Thibado

*Department of Physics, The University of Arkansas, Fayetteville, Arkansas 72701*

The reconstructions of the InP(001) surface prepared by molecular beam epitaxy have been studied with *in situ* reflection high-energy electron diffraction (RHEED) and scanning tunneling microscopy (STM). The growth chamber contains a highly accurate temperature measurement system and uses a solid-source, cracked phosphorus, valved effusion cell. Five InP(001) reconstructions are observed with RHEED by analyzing patterns in three principal directions. Under a fixed $P_2$ flux, decreasing the substrate temperature gives the following reconstructions: $c(2\times8)$, $(2\times4)$, $(2\times1)$, $(2\times2)$, and $c(4\times4)$. *In situ* STM images reveal that only two of these reconstructions yields long-range periodicity in real space. InP(001) does not form the metal rich $(4\times2)$ reconstruction, which is surprising because the $(4\times2)$ reconstruction has been coined the universal surface reconstruction since all III–V(001) surfaces were thought to favor its formation.

I. INTRODUCTION

Indium phosphide (InP) is a technologically important member of the III–V, or compound semiconductor family of materials that are used to make high-speed and optoelectronic devices.[1] Unlike Si-based devices which are primarily formed by ion implantation methods,[2] III–V structures must be formed by depositing one plane of atoms on top of another until the entire device structure is formed. Naturally, surface structure plays an important role in the growth and possibly plays a role in the overall properties of these devices. For example, a certain surface reconstruction may produce low quality crystal growth due to its symmetry properties altering diffusion or nucleation rates. In addition, stoichiometry changes on a surface may produce a nonuniform interface which may have a significant impact on short-period heterostructures where the interfaces constitute a large fraction of the total heterostructure. Therefore, there is a need to better understand III–V(001) surface reconstructions.

To date, the most intensely studied compound semiconductor surface has been the GaAs(001) surface.[3–5] The InP(001) surface reconstructions have received less attention. Like GaAs there are three dominant techniques for preparing the InP surface: sputter-and-anneal, growth of InP using gas-source phosphorous, and growth using solid-source phosphorous. Using gas-source phosphorous, electron diffraction studies have reported a $(2\times2)$, $(2\times1)$, and $(2\times4)$ reconstructions with increasing substrate temperature.[6,7] In these studies the $(2\times2)$ reconstruction is not distinguished from the $c(4\times4)$ since diffraction data is not reported in the [100] direction. Using solid-source phosphorous, electron diffraction studies have observed a $c(4\times4)$, $(2\times2)$, $(2\times1)$, and $(2\times4)$ reconstructions as a function of substrate temperature and phosphorous flux.[8] For these various phases, local structural information is not reported. A $c(4\times4)$ and several $(2\times4)$ reconstructions have been examined theoretically and found to be stable.[9–11] The $(2\times1)$ structure has not been theoretically modeled and one would expect it to be energetically unfavorable because it should violate the electron counting model.[12] To measure the local structure, several scanning tunneling microscopy (STM) experiments have been carried out on various InP(001) surface reconstructions. To date, STM images have not been reported for the $c(4\times4)$ structure. STM studies have observed an ordered $(2\times1)$ reconstruction[13] when prepared using gas-source phosphorous. Other gas-source studies of this surface have found it to be a mixture of several different reconstructions.[14] STM experiments have also observed a surface with a $(2\times4)$ symmetry.[15–18] Still needed is a systematic mapping of the reconstructions versus absolute substrate temperature and solid source phosphorus flux using both electron diffraction and STM.

In this study, all possible surface reconstructions of molecular beam epitaxy (MBE) prepared InP(001) are mapped out using *in situ* reflection high-energy electron diffraction (RHEED) as a function of substrate temperature and solid-source $P_2$ flux, including zero $P_2$ flux. A highly accurate noncontact temperature measurement system is used to measure the absolute temperature of the substrate. *In situ* STM studies provide images of the $c(4\times4)$, $(2\times1)$, and $(2\times4)/c(2\times8)$ for this system and show that only two of these reconstructions have long-range order in real space.

## II. EXPERIMENT

Experiments were carried out in an ultrahigh vacuum (UHV) multi-chamber facility (5–8×10$^{-11}$ Torr throughout) which contains a solid-source MBE chamber (Riber 32P) that includes a substrate temperature determination system accurate to +/- 2 °C.[19] This system also contains a solid source, cracked phosphorus cell with a valved controlled flux. In addition, this chamber is connected to a surface analysis chamber with an STM (Omicron).[20]

For the RHEED measurements, commercially available, "epiready," $n$-type (S doped 10$^{18}$/cm$^3$) 2 in. InP(001) +/- 0.05° substrates were loaded into the MBE system without any chemical cleaning. The surface oxide layer was removed at 490 °C while exposing the surface to a 10 µTorr P$_2$ flux using a cracker temperature of 950 °C. A 1.5-µm-thick InP buffer layer was grown at 460 °C using a growth rate of 1.0 µm/h as determined by RHEED oscillations, and a P$_2$ to In beam equivalent pressure (BEP) ratio of 15. Growth using a cracker results in a film that is unintentionally doped to about 10$^{16}$/cm$^3$ $n$ type.[21] Surface reconstructions for a fixed P$_2$ flux were identified by either heating or cooling the substrate in 10 °C increments, waiting 15 min, and recording the RHEED pattern in the [110], [1-10], and [100] directions. The symmetry of the surface at each temperature was then identified by analyzing the three RHEED patterns. This procedure was repeated for five P$_2$ fluxes by adjusting the valve position on the phosphorus cell. In addition, the surface reconstruction phases were measured without any P$_2$ flux by first creating the $c$(4×4) pattern at low temperatures and low P$_2$ fluxes. Then, the P$_2$ flux was eliminated by closing the valve and waiting 30 min to allow the background P$_2$ to be removed from the chamber by the ion pump. Finally, the RHEED patterns were recorded as described above by heating the substrate in 10 °C increments. For the highest P$_2$ flux and the zero P$_2$ flux data series, the substrate temperature was increased to a temperature where the surface was irreversibly damaged. Attempts were also made to produce the (4×2) reconstruction without success, which included monitoring the RHEED pattern during growth while simultaneously lowering the P$_2$ to In flux ratio.

For the STM measurements, identical substrates were used. The oxide was removed and a buffer layer was grown in the same manner as the RHEED sample. Between the various STM studies, InP was regrown on the substrate at 465 °C with a P$_2$ BEP of 2 µTorr for 15 min using a growth rate of 0.2 ML/s. The substrate was then annealed at 550 °C for 15 min with P$_2$ flux of 13 µTorr followed by another anneal at 480 °C with no P$_2$ flux for 15 min. In order to prepare a particular surface reconstruction with as much long-range order as possible, the sample was annealed under the highest possible P$_2$ flux and temperature that produced that reconstruction for as long as 1 h. This enables the highest atom diffusion rates, thus forming the longest-range order on the surface. After this anneal, the P$_2$ flux was ramped to zero at the same time the substrate temperature was ramped to the highest value that still produces the same reconstruction pattern, but with no P$_2$ flux. During the decrease in temperature the RHEED pattern was monitored to ensure that it remained unchanged. The sample was then annealed for another 30 min without a P$_2$ flux, after which the sample was cooled to room temperature, transferred to the STM without breaking UHV, and imaged at room temperature. For each sample, multiple filled-state STM images were acquired using tips made from single crystal <111>-oriented tungsten wire, a sample bias of -3.0 V and a demanded tunneling current of 0.05–0.2 nA. All STM images have a (001) plane subtracted from the data.

## III. RESULTS

### A. RHEED measurements

The structural transitions between various surface reconstructions as observed by RHEED for InP(001) as a function of P$_2$ flux and substrate temperature are shown in Fig. 1. Increasing the substrate temperature at any nonzero P$_2$ flux results in the surface reconstruction changing from $c$(4×4) to (2×2), (2×1), (2×4), and finally to $c$(2×8). Decreasing the substrate temperature reverses this reconstruction sequence at the same temperatures. The zero P$_2$ pressure data series is shown on a separate plot directly below the logarithmic scale, where the transitions only happen when the substrate temperature is increased starting from the $c$(4×4) phase. For example, the $c$(2×8) phase will remain as the substrate temperature is decreased from 450 °C to room temperature under zero P$_2$ pressure. To the right of the thick dashed line the surface reconstruction remains the $c$(2×8) symmetry. However, the pattern does become dimmer in time and surface degradation is visibly apparent (i.e., large fractions of the surface are cloudy). This condition has been observed in previous studies and to our knowledge it is impossible to recover the original surface morphology once this point is passed.[15]

B. STM measurements

From the RHEED phase diagram displayed in Fig. 1 only the $c(2\times8)/(2\times4)$, $c(4\times4)$, and the $(2\times1)$ reconstructions were imaged with STM. Characteristic STM images of the InP(001) surface after preparing the $c(2\times8)$ surface reconstruction are shown in Fig. 2. A typical large-scale STM image is shown in Fig. 2(a). Here each gray level represents a terrace which is separated from the next by a monolayer high step (0.29 nm). The surface tends to favor steps that run along the [1-10] direction without having kinks. In addition, these steps tend to bunch together as shown in the upper right corner of Fig. 2(a). A higher-magnification image is shown in Fig. 2(b), which shows rows running along the direction. These rows are separated from each other by about 1.7 nm and represent the 4-by periodicity of the surface. At this magnification the surface favors some small pit formation; inside these pits the next layer is visible, which also shows the 4-by rows. An even higher magnification image of the surface is shown in Fig. 2(c). At this scale another periodicity running along the rows is observed. The spacing between these features is about 0.8 nm and represents the 2-by periodicity. Most of these rows have the 2-by periodicity aligned with each other, making the surface appear $(2\times4)$-like, as indicated by the unit-cell box and label. One region near the bottom shows the $c(2\times8)$ periodicity and is also shown with a unit-cell box and label. Notice that there appears to be some variation in the height of the image shown in Fig. 2(c). The exact cause of this effect is unknown but may be due to some local defects or buried dopants beneath the surface.

Characteristic STM images of the InP(001) surface after preparing the $c(4\times4)$ surface reconstruction are shown in Fig. 3. A typical large-scale STM image is shown in Fig. 3(a). Again, each gray level represents a terrace that is separated from the next by a monolayer high step. Notice that this surface seems to favor steps that are rounded in the (001) plane and necessarily have a large kink density. A higher-magnification image is shown in Fig. 3(b), which shows a brick wall-like pattern. Even though the surface has fairly large regions that are well ordered, the surface is not nearly as well ordered as the $c(2\times8)$ surface. The $c(4\times4)$ surface does not have large defect free terraces, rather it frequently has small adatom and vacancy islands on each terrace. An even higher magnification image of the surface is shown in Fig. 3(c). Here the origin or the $c(4\times4)$ surface reconstruction is more clear. The box overlayed on Fig. 3(c) highlights a conventional unit cell. This structure can be described as an atomic-scale brick wall, with staggered rows of bricks running along the [1-10] direction.

Characteristic STM images of the InP(001) surface after preparing the $(2\times1)$ surface reconstruction are shown in Fig. 4. A typical large-scale STM image is shown in Fig. 4(a). It is immediately clear that this surface does not have long-range periodicity. Each terrace is broken up into a large number of islands. The underlying terrace structure is still visible and shows a tendency to favor straight edges running along the [1-10] direction. A higher-magnification image is shown in Fig. 4(b), which shows only small regions with periodic structure. Throughout this image, some small regions show rows running along the [1-10] direction, which is the same direction as the 4-by rows of the $c(2\times8)$ surface reconstruction. However, these rows extend only short distances and then defects occur. There are also long vacancy islands running in this direction, giving the surface several monolayers of roughness. An even higher magnification image of the surface is shown in Fig. 4(c). In the upper right corner of the image a few rows running along the [1-10] direction are indicated by arrows. However, a large number of other complex structures are also present. No $(2\times1)$ unit cell is identifiable.

IV. DISCUSSION

It is insightful to compare and contrast the GaAs(001) surface phase diagram with InP(001) since the GaAs(001) surface is the most widely studied III–V surface.[3] The InP(001) surface exhibits a $c(4\times4)$ reconstruction like GaAs(001). This unique brick wall-like structure is identical to the $c(4\times4)$ surface reconstruction observed on a majority of other III–V systems, such as AlSb, InSb, GaAs, AlAs, and InAs.[22] Most likely, the structural model of the InP(001)-$c(4\times4)$ is the same (i.e., 1.75 planes of P on top of a full plane of In). This structure has been theoretically modeled as the lowest energy anion-rich surface reconstruction.[11] Unlike the GaAs(001) surface there is a large temperature range between the $c(4\times4)$ and $(2\times4)$ reconstructions where the surface is either $(2\times2)$ or $(2\times1)$. Even though the RHEED pattern shows a wide temperature–pressure range where the $(2\times1)$ is favored, the local real-space picture indicates this phase is locally disordered without any identifiable unit cell. This is not surprising since a $(2\times1)$ reconstruction would violate the electron counting model.[12] It is concluded that this phase is simply a disordered transition from the $c(2\times8)$ to the $c(4\times4)$ surface reconstruction. A gas-source STM study of the InP(001) surface by Li *et al.* has observed an ordered $(2\times1)$ structure similar to Si.[13] It is unknown why they observe an ordered structure and we do not. However, high concentrations of other elements, such as hydrogen, are present in their studies and this may be the reason for the difference.

The InP(001) surface exhibits a distinct $c(2\times8)$ RHEED pattern, while GaAs only forms the $(2\times4)$ or a weak $c(2\times8)$. This is most likely a result of a change in the topmost layer of atoms, since the RHEED probe is highly surface sensitive. As previously indicated in Fig. 2(c) a top layer local $c(2\times8)$ symmetry is visible which is not the case for GaAs.[5] The images presented here do not resolve the atomic structure, however, several structural models have been proposed for the $(2\times4)$ symmetry.[9] One model is the mixed dimer $(2\times4)$, which has an In and P atom in the top layer. If these atoms arrange themselves into a $c(2\times8)$ structure, this could be another reason for seeing the $c(2\times8)$ symmetry in RHEED.

Another interesting difference between the GaAs(001) and InP(001) surfaces, is that the GaAs(001) surface exhibits a unique $(4\times6)$ reconstruction *only* when heated with no $As_4$ flux incident on the surface.[23] There is no unique reconstruction that appears for the InP(001) surface when there is no $P_2$ present.

The most surprising difference we found was that all attempts to make a $(4\times2)$ reconstruction appear on the InP(001) surface were unsuccessful. Not only does this reconstruction occur on the GaAs(001) surface, it has also been observed on all other III–V(001) surfaces and was thought to be a universal reconstruction.[24] The inability to create a $(4\times2)$ reconstruction is in good agreement with recent theoretical studies, which indicate that the cation-rich (In-rich) InP(001) surface favors the formation of a mixed-dimer $(2\times4)$ reconstruction over the $(4\times2)$.[25–27] In fact, this reconstruction has been observed experimentally with STM and confirmed theoretically.[15,27] Theoretical studies show that at low temperatures the $\beta2(2\times4)$ is favored, while at high temperature the mixed dimer $(2\times4)$ is favored. It is possible the phase transition from $(2\times4)$ to $c(2\times8)$ that we report is coincident with this structural change. Finally, if the InP(001) surface is heated high enough it was irreversibly damaged. This is unlike the GaAs(001) surface, and is most likely due to In droplet formation making the surface metallic.

## V. CONCLUSION

All the reconstruction phases of the InP(001) surface prepared by solid source MBE have been mapped out as a function of $P_2$ flux and temperature with *in situ* RHEED. Under fixed $P_2$ flux, five InP(001) surface reconstructions are observed with increasing temperature: $c(4\times4)$, $(2\times2)$, $(2\times1)$, $(2\times4)$, and $c(2\times8)$. The surface irreversibly degrades on optical length scales when heated too high. The local order was investigated with STM and only the $c(4\times4)$ and $(2\times4)/c(2\times8)$ have an identifiable unit cell. The behavior of the InP(001) surface was found to have many differences from the GaAs(001) surface.

ACKNOWLEDGMENTS

This work was supported by the Office of Naval Research (ONR) under Grant No. N00014-97-1-1058 and the National Science Foundation under Grant No. DMR-9733994.

FIG. 1. RHEED-derived surface reconstruction transition temperatures for InP(001) as a function of incident $P_2$ BEP. The zero $P_2$ pressure phases are not shown on the logarithmic scale, but are shown on the lower graph. The solid lines represent least-squares fits to the data points shown as filled squares. All phases shown are reversible, except for the zero $P_2$ pressure, which are only applicable for increasing the temperature starting from the $c(4\times4)$ phase. Across the thick dashed line the $c(2\times8)$ phase remains, however the onset of optically visible and irreversible surface degradation occurs.

FIG. 2. STM images for the InP(001)-$(2\times4)$ surface reconstruction: (a) 1000 nm × 1000 nm STM image showing flat terraces and step bunching; (b) 100 nm×100 nm STM image showing the 4-by periodicity in the [110] direction; (c) 20 nm×20 nm STM image showing the 4-by and 2-by periodicity in the [110] and [1-10] directions, respectively. In addition, the $(2\times4)$ and $c(2\times8)$ unit cells are drawn over their corresponding regions.

FIG. 3. STM images for the InP(001)-$c(4\times4)$ surface reconstruction: (a) 1000 nm × 1000 nm STM image showing flat terraces with rounded step edges in the (001) plane; (b) 100 nm×100 nm STM image showing regions that are $c(4\times4)$ and regions that are disordered; (c) 10 nm×10 nm STM image showing the brick wall-like structure of $c(4\times4)$ reconstruction. A conventional unit cell is draw over the image in (c).

FIG. 4. STM images for the InP(001)-$(2\times1)$ surface reconstruction: (a) 1000 nm × 1000 nm STM image showing a surface with several monolayers of roughness and step edges running along the [1-10] direction; (b) 100 nm × 100 nm STM image showing regions with several monolayers of roughness and regions with row-like structures running along the [1-10] direction; (c) 10 nm×10 nm STM image showing arrows highlighting the [1-10] rows and that a unit cell is not identifiable.

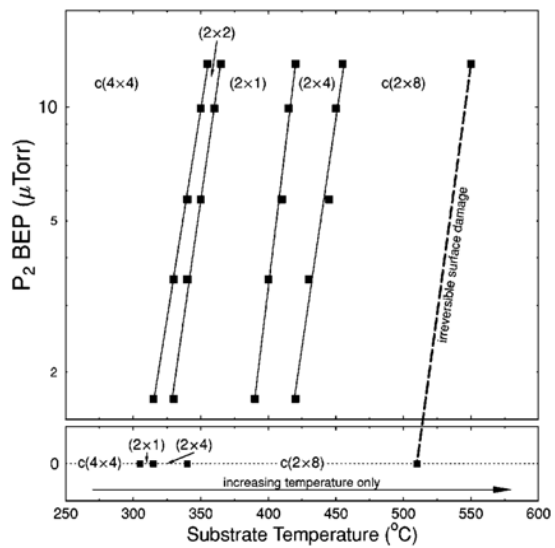

Figure 1.

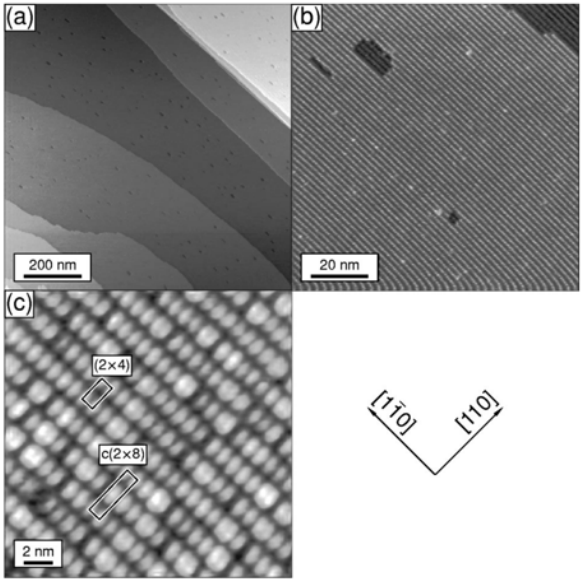

Figure 2.

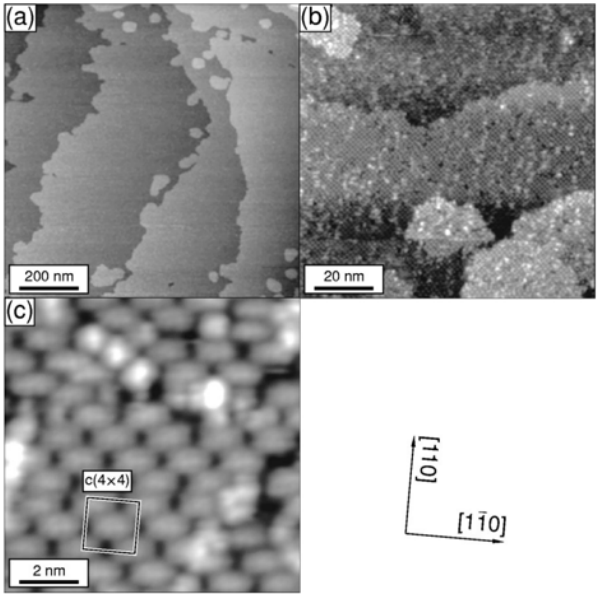

Figure 3.

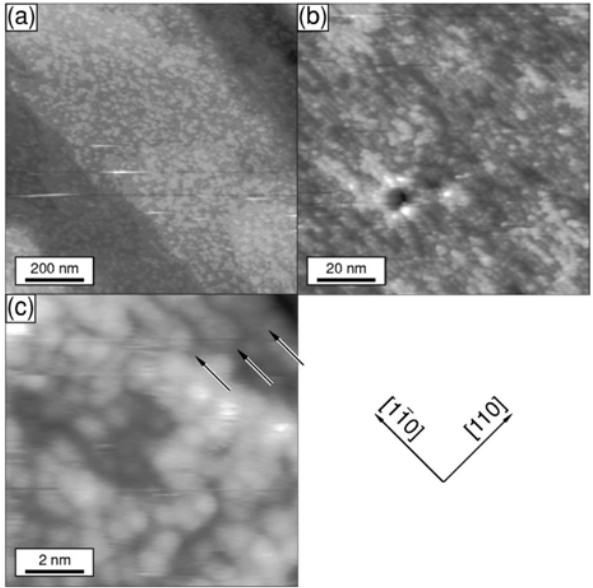

Figure 4.